\documentclass[aps,pre,reprint,superscriptaddress,letterpaper]{revtex4-1}

\usepackage{amssymb}
\usepackage{amsmath}
\usepackage{bm}

\usepackage{graphicx}
\graphicspath{{fig/}}
\usepackage{grffile}
\usepackage{subfigure}

\usepackage{hyperref}
\hypersetup{
  hidelinks,
}

\usepackage{cleveref}

\begin{document}

\title{DCA for genome-wide epistasis analysis: the statistical genetics perspective}

\author{Chen-Yi Gao}
\email{gaocy@itp.ac.cn}
\affiliation{Key Laboratory of Theoretical Physics,
  Institute of Theoretical Physics, Chinese Academy of Sciences,
  Beijing 100190, China}

\author{Fabio Cecconi}
\email{fabio.cecconi@cnr.it}
\affiliation{CNR-ISC and Dipartimento di Fisica, Sapienza Universit\'a di Roma, p.le A. Moro 2, 00185 Roma, Italy}

\author{Angelo Vulpiani}
\email{angelo.vulpiani@roma1.infn.it}
\affiliation{CNR-ISC and Dipartimento di Fisica, Sapienza Universit\'a di Roma, p.le A. Moro 2, 00185 Roma, Italy}
\affiliation{Centro Interdisciplinare B. Segre, Accademia dei Lincei, Roma, Italy}

\author{Hai-Jun Zhou}
\email{zhouhj@itp.ac.cn}
\affiliation{Key Laboratory of Theoretical Physics,
  Institute of Theoretical Physics, Chinese Academy of Sciences,
  Beijing 100190, China}
\affiliation{School of Physical Sciences,
  University of Chinese Academy of Sciences, Beijing 100049, China}

\author{Erik Aurell}
\email{eaurell@kth.se}
\affiliation{Department of Computational Biology,
  KTH-Royal Institute of Technology, SE-10044 Stockholm, Sweden}
\affiliation{Depts of Applied Physics and Computer Science, Aalto University,
  FIN-00076 Aalto, Finland}

\begin{abstract}
Direct Coupling Analysis (DCA)
is a now widely used method to leverage statistical information 
from many similar biological systems to draw meaningful
conclusions on each system separately.
DCA has been applied with great success
to sequences of homologous proteins, and also more recently 
to whole-genome population-wide sequencing data.
We here argue that the use of DCA
on the genome scale is contingent on fundamental
issues of population genetics. DCA can be expected to
yield meaningful results when a population is in
the Quasi-Linkage Equilibrium (QLE) phase studied by
Kimura and others, but not, for instance, in a phase of Clonal
Competition. We discuss how the exponential (Potts model)
distributions emerge in QLE,
and compare couplings to correlations
obtained in a study of about 3,000 genomes of the human pathogen 
\textit{Streptococcus pneumoniae}.
\end{abstract}

\date{\today}

\maketitle

\section{Introduction}
\label{sec:introduction}
Direct Coupling Analysis (DCA) is a collective term for a number
of related techniques to learn the parameters in Ising/Potts models
from data and to use these inferred parameters in biological data analysis~\cite{Nguyen-2017a}.
DCA has led to a breakthrough in identifying
epistatically linked sites in proteins from protein sequence data~\cite{Weigt-2009a,Morcos-2011a,Stein-2015a,Michel-2017a,Cocco-2018a},
which in turn has been used to predict spatial contacts from the sequence data~\cite{Ovchinnikov-2017a,Michel-2017b,Ovchinnikov-2018a}.
DCA has also been used to identify 
nucleotide-nucleotide contacts of RNAs~\cite{DeLeonardis-2015a},
multiple-scale protein-protein interactions~\cite{Gueudre-2016a,Uguzzoni-2017a},
amino acid--nucleotide interaction in RNA-protein complexes~\cite{Weinreb-2016a}
and synergistic effects not necessarily related to spatial contacts~\cite{Figliuzzi-2016a,Hopf-2017a,CouceE9026,Schubert325993}.

Skwark \textit{et al} applied a version of DCA to
whole-genome sequencing data of a population of \textit{Streptoccoccus pneumoniae}~\cite{Skwark-2017a},
and were able to retrieve interactions
between members of the Penicillin-Binding Protein (PBP) family
of proteins as well as other predictions.
\textit{S. pneumoniae} (pneumococcus) is an important human pathogen 
where resistance to antibiotics in the $\beta$-lactam family of compounds
are associated to alterations in their target enzymes, which are the PBPs~\cite{Hakenbeck2012}.
Further results were recently given in~\cite{Puranen-2017a}
showing robustness by using sequencing data from a 
pneumococcal population from another continent, 
and identifying a novel seasonal phenotype signal.
Three of the authors of the current article additionally recently
showed that DCA analysis on the bacterial genome scale does not
need supercomputing resources, but can be carried out
in a reasonable time (hours) on a standard desktop computer~\cite{GaoZhouAurell2018}.    

These advances raise the question why DCA works at all,
and if one can identify from the outset when that is the case.
As discussed by one of us ``max-entropy'' arguments sometimes
evoked in the literature are not pertinent to this issue~\cite{Aurell2016}.
Instead, we will here argue that at least for genome-scale data
the answer lies in a very different direction.
We will show that the \textit{Quasi-Linkage Equilibrium} (QLE) 
of Kimura~\cite{Kimura1956,Kimura1964,Kimura1965},
as extended by Neher and Shraiman to statistical genetics on the genome scale~\cite{NeherShraiman2009,NeherShraiman2011},
provides a natural and rational basis for DCA. 
According to this theory a population evolving 
with sufficient amount of exchange of genetic material
(recombination, or any form of sex) will settle down
to a dynamic equilibrium where the 
distribution of genotypes is of the form assumed by DCA.
In the opposite case of little exchange of genetic material
(little sex) the distribution over genotypes
is different and dominated by clones, 
identical or very similar individuals
descended from a common ancestors.
In such a setting DCA is not an appropriate approach, and is likely to yield nonsense results.

We will also discuss the inference task of DCA in the context of QLE
as realistically applied to biological data.
We will first show that DCA can give a much more sparse
representation of the data than correlations (covariances).
This is in line with the intended meaning of the acronym DCA: 
the parameters in a Potts or an Ising model can be considered
``direct couplings'', and while these are typically reflected in correlations (covariances),
the latter also includes combined effects, or ``indirect couplings''.
Second, the authors of~\cite{NeherShraiman2009,NeherShraiman2011}
assumed that a genotype can be described by a Boolean vector
\textit{i.e.} a string of $0$'s and $1$'s. This is almost never
the case for population-wide whole-genome sequencing data
due to varying gene content, which have to be represented as gaps.
We have therefore generalized the theory
to categorical data and to a model of bacterial recombination.
Third, as surveyed in~\cite{Nguyen-2017a}, DCA as a methodology has matured 
considerably
over the last decade.
For the mathematical task of inferring parameters in a Potts or Ising model   
from data which was generated from such a model, the 
Small-Interaction Expansion (SIE) used in~\cite{NeherShraiman2009,NeherShraiman2011}
is inferior to many other inference methods.
We will show that it is also inferior when applied to
real data in the sense of yielding much less sparse
results, and would also have specific problems when applied to simulation data. 
A conclusion of this work is hence that when
QLE is combined with DCA on the genome scale, it should be combined
with one of the modern and more powerful versions of DCA.

The paper is organized as follows.
Sections~\ref{sec:QLE}-\ref{sec:log-Prob}
reformulate the theory of \cite{NeherShraiman2011}
in a way suitable to our presentation and for categorical data.
Section~\ref{sec:QLE} hence contains a non-mathematical
overview, while Sections~\ref{sec:Neher-Shraiman} and~\ref{sec:modifications} 
contain the specific changes needed for categorical data and our model
of bacterial recombination.
Section~\ref{sec:log-Prob} formulates the dynamics of
Potts model parameters in QLE phase, which is a central result of the theory.
Section~\ref{sec:DCA-sparse} presents results for 
real sequence data and Section~\ref{sec:QLE-genetic-drift} for simulation data.
Section~\ref{sec:discussion} contains discussion and outlook for future work.
Technical details such as 
a derivation of SIE for categorical data,
sequence and code availability are given in appendices.

\section{Statistical genetics and Quasi-Linkage Equilibrium}
\label{sec:QLE}
We will here first present key concepts and results in a mostly
non-mathematical manner.
The driving forces of evolution are assumed to be genetic drift, 
mutations, recombination, and fitness variations. The first refers to the element of chance; in a finite 
population it is not certain which genotypes will reproduce and leave
descendants in later generations. The three latter are deterministic,
describing the expected success or failure of different
genotypes. Mutations are hence random genome changes described by mean rates. 

Recombination (or sex) is the mixing of genetic material between different
individuals.
In diploid organisms every individual inherits 
half of its genetic material
from the mother, and half from the father. This material is further
mixed up in the process called cross-over so that each chromosome
of the child consists of segments alternately inherited from the mother and the father.
By sequencing the parents and children in a single family
the per generation mutation rate and number of cross-over segments in human 
has been measured to be about $30$ and $100$~\cite{Roach636},
numbers that are in line with previous estimates.
By this measure recombination is hence in human about three times faster than mutations.
In bacteria recombination happens by transformation, transduction, and conjugation.
The ratio of recombination to mutations 
differ greatly between different bacterial species 
and can also differ between different strains and different environments of the same species.  
In this work we use data from \textit{S. pneumoniae}
where this ratio has been estimated from less than one to over 
forty, but with an average close to nine~\cite{Chaguza2016}.
Similarly to the analysis in~\cite{NeherShraiman2011}
we will for the most part here assume that
recombination is a faster and stronger effect than mutations.

Fitness means in statistical genetics a propensity 
for a given genotype to propagate its genomic material to the next generation.
Like mutation and recombination fitness is hence here a rate, measured in units $\left(\hbox{time}\right)^{-1}$.
Fitness variations refer to the variations of these rates.
Consider then then the effects of recombination and fitness
on correlated variations in a population, ignoring mutations and
genetic drift. The correlation between alleles $\alpha$ and
$\beta$ at loci $i$ and $j$ is $M_{ij}(\alpha,\beta)=f_{ij}(\alpha,\beta)
- f_{i}(\alpha)f_{j}(\beta)$ where $f_{i}(\alpha)$ is the frequency
of allele $\alpha$ at locus $i$ and similarly $f_{j}(\beta)$,
and where $f_{ij}(\alpha,\beta)$ is the frequency of simultaneously finding
alleles $\alpha$ and $\beta$ at loci $i$ and $j$. 
If there is recombination between $i$ and $j$
but 
are no fitness variations at all,
then it is trivial to see that $M_{ij}(\alpha,\beta)$ must decay to zero.
This state is called \textit{Linkage Equilibrium} (LE).

If now instead fitness variations are small but non-zero,
then non-zero correlations
may persist.
We will assume that the fitness of genotype $\mathbf{g}$ which carries allele $g_i$ on locus $i$ depends on single-locus
variations and pair-wise co-variations, that is
\begin{equation}
  \label{eq:fitness}
  F(\mathbf{g}) = F_0 + \sum_i F_i(g_i) + \sum_{ij}F_{ij}(g_{i},g_{j})
\end{equation}
If so, the first central result of statistical genetics is that when recombination is
sufficiently strong, the distribution will have the form
\begin{equation}
  \label{eq:Potts}
P(\mathbf{g}) = \frac{1}{Z}\exp\left(\sum_i h_i(g_i) + \sum_{ij} J_{ij}(g_i,g_j)\right) 
\end{equation}
The above distribution is also the Gibbs-Boltzmann distribution over variables $\mathbf{g}$
with energy terms $h_i$ and $J_{ij}$, and where $Z$ (the partition function) is the normalization.
The second central result is that
\begin{equation}
  \label{eq:main-Neher-Shraiman-eq}
  J_{ij}(\alpha,\beta) = \frac{F_{ij}(\alpha,\beta)}{rc_{ij}}
\end{equation}
where $r$ is an overall recombination rate and $c_{ij}$ is the probability that alleles at loci $i$ and $j$ are inherited from the
same parent.  
For the most part we will in the following assume that $c_{ij}$ equals $\frac{1}{2}$, appropriate if
recombination is sufficiently strong and 
loci $i$ and $j$ are sufficiently far apart on the genome.
Note that the right-hand
side of \eqref{eq:main-Neher-Shraiman-eq} is the ratio of two rates, and therefore dimension-less.
For the distribution \eqref{eq:Potts} the parameters
$J_{ij}(\alpha,\beta)$ carry the same information as the correlations $M_{ij}(\alpha,\beta)$ but in a de-convoluted or ``direct'' manner.

Inferring $J_{ij}(\alpha,\beta)$ from data is what 
the methods known as DCA achieve~\cite{Nguyen-2017a}.
From \eqref{eq:main-Neher-Shraiman-eq} this gives
fitness parameters $F_{ij}(\alpha,\beta)$ up to a proportionality (the overall rate $r$),
and for pairs of loci sufficiently far apart on the genome so that $c_{ij}$ is approximately constant.
Recombination does not change single-locus frequencies; in a stationary state parameters $h_i(\alpha)$ instead result from a dynamic equilibrium between
fitness and mutations. In the absence of mutations QLE in an infinite population
is in fact only a long-lived transient while the $h_i(\alpha)$ change slowly in time
as the population drifts towards fixation. 
In a finite population both $h_i(\alpha)$ and
$J_{ij}(\alpha,\beta)$
also fluctuate in time, and the prediction \eqref{eq:main-Neher-Shraiman-eq} does not apply directly.
All these aspects have to be taken into account when applying DCA techniques to 
analyze a QLE phase.

The third central prediction of statistical genetics is that when fitness variations still have the form \eqref{eq:fitness} but
are not small compared to recombination, then the distributions
will not be of the form \eqref{eq:Potts}. In that phase, in~\cite{NeherShraiman2009,NeherShraiman2011}
called \textit{Clonal Competition} (CC), the distribution is instead better described
as
\begin{equation}
  \label{eq:mixture-model}
  P(\mathbf{g}) = \sum_c \mu_c P_c(\mathbf{g})
\end{equation}
where the sum goes over clones, $\mu_c$ is the weight of clone $c$, and $P_c(\mathbf{g})$ is some distribution peaked
around clone center $\mathbf{g}_c$.
Statistical genetics hence predicts a parameter-dependent transition between the two canonical distribution families
in high-dimensional statistics, namely the exponential model (\eqref{eq:Potts}) and the mixture model
(\eqref{eq:mixture-model}). 
A further difference between QLE and CC is that in QLE the joint distribution over more than
one genotype approximately factorizes, $P(\mathbf{g}_1,\ldots,\mathbf{g}_N)\approx P(\mathbf{g}_1)\cdots P(\mathbf{g}_N)$.
In CC phase this is not so; genomes related by descent do not vary independently. 
A different analogy, discussed in~\cite{NeherShraiman2009,NeherShraiman2011}
is that of equilibrium states in disordered
systems~\cite{MezardParisiVirasoro,MezardMontanari};
\eqref{eq:Potts} is like a high-temperature replica-symmetric para-magnetic phase, while \eqref{eq:mixture-model} is like a low-temperature replica symmetry-breaking
spin glass phase. 

\section{Statistical genetics for categorical data}
\label{sec:Neher-Shraiman}
In this section we summarize statistical genetics as formulated in~\cite{NeherShraiman2009,NeherShraiman2011}
in a more technical manner, and generalize the theory to categorical data \textit{i.e.} to
when there can be more than two alleles per locus.
Let there be $M_i$ alleles at locus $i$ and
let the allele be indicated by a variable $l_i$ that takes values $1,2,\ldots,M_i$.
The frequency of allele $\alpha$ at locus $i$ is
\begin{equation}
  f_i(\alpha) = \left< \mathbf{1}_{l_i,\alpha} \right>
\end{equation}
These quantities satisfy $\sum_{\alpha=1}^{M_i} f_i(\alpha) = 1$.
The covariance matrix between loci $i$ and $j$ is
\begin{equation}
    \label{eq:c-ij}
  M_{ij}(\alpha,\beta) = \left< \mathbf{1}_{l_i,\alpha} \mathbf{1}_{l_j,\beta}\right> - f_i(\alpha) f_j(\beta) 
\end{equation}
These quantities satisfy $\sum_{\alpha=1}^{M_i}M_{ij} (\alpha,\beta) = \sum_{\beta=1}^{M_j}M_{ij}(\alpha,\beta) = 0$.
The variance matrix at one locus is
\begin{equation}
  \label{eq:c-ii}
  M_{ii}(\alpha,\beta) = \mathbf{1}_{\alpha,\beta} f_i(\alpha)  - f_i(\alpha) f_j(\beta)  
\end{equation}
and satisfies $\sum_{\alpha=1}^{M_i}M_{ii} (\alpha,\beta) = \sum_{\beta=1}^{M_j}M_{ii}(\alpha,\beta) = 0$.

Statistical genetics are evolution equations for the distributions over genotypes
\begin{equation}
  \frac{d}{dt} P(\mathbf{g}) =   \frac{d}{dt}|_{mut} P(\mathbf{g}) + \frac{d}{dt}|_{fitness} P(\mathbf{g}) + \frac{d}{dt}|_{recomb} P(\mathbf{g}) 
\end{equation}
where the three terms on the right-hand side represent the changes due to mutations, fitness variations and recombination.
The mechanisms of mutations and fitness are classical in population genetics, and known as Wright-Fisher models.

Single-locus mutations are hence modelled by matrices $\mu^{(i)}_{\alpha, \beta}$
which give the rate at which allele $\alpha$ on locus $i$ changes to allele $\beta$.
Let $F^{(i)}_{\alpha, \beta}$ be the operator which if the allele at locus $i$ is $\alpha$ changes
it to $\beta$, and otherwise does nothing. In the dynamic equation for probability
mutations hence enter as
\begin{equation}
  \label{eq:evolution-mutations}
  \frac{d}{dt}|_{\hbox{mut}} P \left(\mathbf{g} \right)= \sum_i \sum_{\alpha\beta} \mathbf{1}_{g_i,\alpha}\left(
  \mu^{(i)}_{\beta,\alpha} P \left( F^{(i)}_{\alpha,\beta} \mathbf{g}\right) - \mu^{(i)}_{\alpha,\beta}  P \left(\mathbf{g} \right) \right)
\end{equation}
This gives contributions to the dynamic equations for the frequencies and correlations as
\begin{eqnarray}
  \label{eq:f-i-mutations}
  \frac{d}{dt}|_{\hbox{mut}} f_i(\alpha) &=& \sum_{\gamma} \mu^{(i)}_{\gamma,\alpha} f_i(\gamma) - \sum_{\gamma} \mu^{(i)}_{\alpha,\gamma} f_i(\alpha) \\
    \label{eq:c-ij-mutations}
    \frac{d}{dt}|_{\hbox{mut}} M_{ij}(\alpha,\beta) &=& \sum_{\gamma} \mu^{(i)}_{\gamma,\alpha} M_{ij}(\gamma,\beta) - \sum_{\gamma} \mu^{(i)}_{\alpha,\gamma} M_{ij}(\alpha,\beta) + \nonumber \\
                                                && \sum_{\delta} \mu^{(j)}_{\delta,\beta} M_{ij}(\alpha,\delta) - \sum_{\delta} \mu^{(j)}_{\beta,\delta} M_{ij}(\alpha,\beta)  
\end{eqnarray}
In the simulations reported below all transition rates $\mu^{(i)}_{\alpha, \beta}$ are the same.
As discussed above it is often a reasonable assumption to take mutations a weaker effect than
fitness variations and recombination.

Fitness variations, such as \eqref{eq:fitness} above, act on the distributions over genotypes as
\begin{equation}
  \label{eq:evolution-fitness}
  \frac{d}{dt}|_{\hbox{fitness}} P \left(\mathbf{g} \right)= \left(F(\mathbf{g})-\left<F\right>\right) P \left(\mathbf{g} \right)
\end{equation}
where $\left<F\right>=\sum_{\mathbf{g}}F(\mathbf{g}) P \left(\mathbf{g}\right)$ is the instantaneous average of fitness over the population.

Potts models are defined as 
\begin{equation}
  P(\mathbf{g}) = \frac{1}{Z} \exp\left( \sum_{i,\alpha} h_{i}(\alpha) \mathbf{1}_{g_i,\alpha} + \sum_{i,j,\alpha,\beta} J_{ij}(\alpha,\beta) \mathbf{1}_{g_i,\alpha} \mathbf{1}_{g_j,\beta}\right)
\end{equation}
As written the model over-parametrized since the same distribution is found by shifting all
$h_{i}(\alpha)$ by a constant $c_i$ or all $J_{i,j}(\alpha,\beta)$ by a vector $c_{ij}(\beta)$, or a vector $c_{ij}(\alpha)$.
In the DCA literature it is customary to go to the Ising gauge~\cite{Weigt-2009a,Ekeberg-2014a} given by
\begin{equation}
  \sum_{\alpha} h_{i}(\alpha) = \sum_{\alpha} J_{i,j}(\alpha,\beta) = \sum_{\beta} J_{i,j}(\alpha,\beta) = 0
\end{equation}

\section{Bacterial recombination in statistical genetics}
\label{sec:modifications}
Recombination (or sex) takes many different forms depending on if the organism is haploid or diploid
and the type of recombination. The mechanism formulated in~\cite{NeherShraiman2009,NeherShraiman2011}
is specifically for sexual reproduction in haploid yeast, where two parents each produce a mating body (copy of parent genome),
and these two mating bodies merge and produce one new genome while the other half of the genetic material of the two mating bodies is discarded.
As closer to our data we consider instead a form of bacterial recombination,
for which however the evolution essentially turns out to be the same,
modulo a \textit{Stosszahlansatz}.

Recombination is thus (we assume) distinguished by
two genomes merging and forming two new genomes. 
In an elementary step two genotypes are therefore lost (the parents) and
two genotypes are gained (the offspring). Let $E_{\mathbf{g}_1,\mathbf{g}_2 \to \mathbf{g}_1',\mathbf{g}_2'}$ be the event that
two individuals with genotypes $\mathbf{g}_1$ and $\mathbf{g}_2$
recombine and give two individuals $\mathbf{g}_1'$ and $\mathbf{g}_2'$.
To describe the kinetics of the individual process we assume that recombination between the two parents happen with
rate $r Q(\mathbf{g}_1,\mathbf{g}_2)$ where $r$ an overall rate of recombination
and $Q(\mathbf{g}_1,\mathbf{g}_2)$ a relative rate.
The two new genotypes
$\mathbf{g}_1'$ and $\mathbf{g}_2'$ are specified by an indicator variable $\mathbf{\xi}$:
\begin{eqnarray}
  \mathbf{g}_1':\quad   g^{(1)'}_i &=&  \xi_i g_i^{(1)}  + (1-\xi_i) g_i^{(2)} \\
  \mathbf{g}_2':\quad   g^{(2)'}_i &=& (1-\xi_i) g_i^{(1)} + \xi_i g_i^{(2)} 
\end{eqnarray}
and this outcome of the recombination happens with probability $C(\mathbf{\xi})$.
The total rate of the individual event is hence $r Q(\mathbf{g}_1,\mathbf{g}_2)C(\mathbf{\xi})$.
The change of the distribution over genotypes due to recombination is given by
\begin{eqnarray}
  \label{eq:Potts-recomb-bact}
  \frac{d}{dt}|_{rec} P(\mathbf{g}) &=&  r \sum_{\mathbf{\xi},\mathbf{g}'}   C(\mathbf{\xi})
  \big[ Q(\mathbf{g}_1,\mathbf{g}_2) P_2(\mathbf{g}_1,\mathbf{g}_2) - \nonumber \\
         && Q(\mathbf{g},\mathbf{g}')    P_2(\mathbf{g},\mathbf{g}') \big]
\end{eqnarray}
In practice it is hard to use \eqref{eq:Potts-recomb-bact} without assuming that the pair probabilities factorize,
as in gas collisions at low densities in statistical physics.
We assume for simplicity also
that $Q$ depends only on the overlap $q$ between
the two genotypes $\mathbf{g}$ and $\mathbf{g}'$:
\begin{equation}
  q(\mathbf{g},\mathbf{g}') = \frac{1}{L}\sum_{i=1}^L \mathbf{1}_{g_i,g_i'}
\end{equation}
Recombination as modelled above does not change the overlap.
This can be seen as follows:: $q(\mathbf{g}_1,\mathbf{g}_2) = 1 - \frac{1}{L}\sum_{i=1}^L \mathbf{1}_{l_i^{(1)},l_i^{(2)}}$
and $\mathbf{1}_{l_i^{(1)},l_i^{(2)}}=\xi_i(1-\xi_i)\mathbf{1}_{l_i,l_i} + \xi_i^2\mathbf{1}_{l_i,l_i'} + (1-\xi_i)^2\mathbf{1}_{l_i',l_i} + (1-\xi_i)\xi_i \mathbf{1}_{l_i',l_i'}$. As the indicator variable takes values zero and one 
this gives $\mathbf{1}_{l_i^{(1)},l_i^{(2)}}=\mathbf{1}_{l_i,l_i'}$.
   
By this invariance of overlaps 
and the factorization assumption the right-hand side of \eqref{eq:Potts-recomb-bact} simplifies to:
\begin{eqnarray}
  &&r \sum_{\mathbf{\xi},\mathbf{g}'} C(\mathbf{\xi}) P(\mathbf{g}')   Q(\mathbf{g},\mathbf{g}') \sum_{i,j,\alpha,\beta} J_{i,j}(\alpha,\beta) \Big(
  \mathbf{1}_{g_i^{(1)},\alpha} \mathbf{1}_{g_j^{(1)},\beta} + \nonumber \\
  &&\qquad \mathbf{1}_{g_i^{(2)},\alpha} \mathbf{1}_{g_j^{(2)},\beta}- \mathbf{1}_{g_i,\alpha} \mathbf{1}_{g_j,\beta}-\mathbf{1}_{g_i',\alpha} \mathbf{1}_{g_j',\beta}\Big) \nonumber \\
  \label{eq:NS-B4b}
  &=& \sum_{i,j,\alpha,\beta} c_{ij} J_{i,j}(\alpha,\beta) \Big(
                                                        \mathbf{1}_{g_i,\alpha} \hbox{E}_{Q}\left[\mathbf{1}_{g_j',\beta} \right]
                                                       + \hbox{E}_{Q}\left[\mathbf{1}_{g_i',\alpha}\right] \mathbf{1}_{g_j,\beta}
                                                       \nonumber \\
 &&\qquad\qquad                                        - \left<Q\right> \mathbf{1}_{g_i,\alpha}  \mathbf{1}_{g_j,\beta}
                                                       - \hbox{E}_{Q}\left[\mathbf{1}_{g_i',\alpha} \mathbf{1}_{g_j',\beta} \right] \Big)
\end{eqnarray}
where we have used the abbreviations
\begin{eqnarray}
c_{ij} &=&\sum_{\mathbf{\xi}} C(\mathbf{\xi})\left(\xi_i(1-\xi_j)+(1-\xi_i)\xi_j\right) \\
\left<Q\right> &=& \sum_{\mathbf{g}'}Q(\mathbf{g},\mathbf{g}')P(\mathbf{g}') \\
\hbox{E}_Q\left[\mathbf{1}_{g_i',\alpha} \right] &=& \sum_{\mathbf{g}'} \mathbf{1}_{g_i',\alpha} Q(\mathbf{g},\mathbf{g}')P(\mathbf{g}') \\
\hbox{E}_Q\left[\mathbf{1}_{g_i',\alpha}\mathbf{1}_{g_j',\beta} \right] &=& \sum_{\mathbf{g}'} \mathbf{1}_{g_i',\alpha} \mathbf{1}_{g_j',\beta}  Q(\mathbf{g},\mathbf{g}')P(\mathbf{g}')
\end{eqnarray}
The first of these is the probability that two loci are
inherited from the same parent and does not (for this model)
depend on the genotype $\mathbf{g}$.
The last three averages on the other hand depend on $\mathbf{g}$. However, if
the function $Q$ is not too sharply focused the dependence can be taken
weak. In particular, we assume that $\left<Q\right>$ is self-averaging, and essentially does not depend on $\mathbf{g}$. 
In spin glass physics language we hence assume that
$\left<Q\right>$, $\left<Q \mathbf{1}_{l_i',\alpha}\right>$ and $\left<Q\mathbf{1}_{l_i',\alpha}\mathbf{1}_{l_j',\beta}\right>$
are self-averaging in the ``paramagnetic'' phase where QLE is expected to hold.

\section{Evolution equation for log-probability in QLE}
\label{sec:log-Prob}
In QLE the evolution equation can conveniently be written for the logarithmic probability
\begin{equation}
  \label{eq:logPotts}
  \frac{d}{dt}\log P(\mathbf{g})= -\frac{\dot{Z}}{Z} + \sum_{i,\alpha}\dot{h}_{i}(\alpha)
                                  \mathbf{1}_{g_i,\alpha} + \sum_{i,j,\alpha,\beta}\dot{J}_{i,j}(\alpha,\beta) \mathbf{1}_{g_i,\alpha} \mathbf{1}_{g_j,\beta}     
\end{equation}
and the various terms identified. Fitness enters (\ref{eq:logPotts}) as
\begin{equation}
  \frac{d}{dt}|_{fitness} h_{i}(\alpha) = f_{i}(\alpha) \qquad
  \frac{d}{dt}|_{fitness} J_{i,j}(\alpha,\beta) = f_{i,j}(\alpha,\beta)              
\end{equation}
and if there were higher-order terms in fitness (more than pair-wise dependencies)
they would enter higher than quadratic terms in the QLE distribution in the same way.
Ignoring mutations and genetic drift we have for the pair-wise dependencies
\begin{equation}
  \label{eq:final-result}
  \dot{J}_{i,j}(\alpha,\beta) = f_{i,j}(\alpha,\beta) -r \left<Q\right> c_{ij}  J_{i,j}(\alpha,\beta)
\end{equation}
where the contribution from recombination is simply read off from \eqref{eq:NS-B4b}.
\eqref{eq:final-result} is a relaxation equation which pushes the Potts model parameter $J_{i,j}(\alpha,\beta)$
to be the ratio of two rates, \eqref{eq:main-Neher-Shraiman-eq} above.
When the data is from one population in a stationary state the average relative rate $\left<Q\right>$
can be subsumed in the overall rate $r$.
When gentoypes are Boolean vectors this gives the same result as Eq.~25 in~\cite{NeherShraiman2011}.

As observed above recombination does not change single-locus frequencies,
and without mutations the $f_{i}(\alpha)$ will drift towards fixation (taking values $0$ or $1$).
Once the population has reached fixation at locus $i$ there can no longer
be any non-zero correlation of Potts parameter involving $i$, and in 
such a setting 
QLE is therefore only a long-lived quasi-stationary state (for the correlations, and for the $J_{i,j}(\alpha,\beta)$'s).
Note that by \eqref{eq:NS-B4b} recombination terms enter in the 
evolution equations of $h_{i}(\alpha)$, in combination with the quantities $J_{i,j}(\alpha,\beta)$.
This is no contradiction, because   
when correlations are non-zero there is not a one-to-one relation between
single-locus frequencies ($f_{i}(\alpha)$) and Potts model magnetization parameters ($h_{i}(\alpha)$);
recombination can influence the latter but not the former.

The break-down of the relaxational equation \eqref{eq:final-result} when the 
single-locus frequencies go to fixation can be understood as follows.
In such a setting the $J_{i,j}(\alpha,\beta)$'s would first remain of order unity, 
while the $h_{i}(\alpha)$'s would tend to $\pm\infty$.
When the $h_{i}(\alpha)$'s become large enough that the minor alleles in a finite-$N$
population are likely to be present only in a few copies, 
a few random events can remove all of the remaining ones at once, 
which sets the correlation and the $J_{i,j}(\alpha,\beta)$ to zero in one go.
Alternatively the argument can be made starting from Eqs.~37-29 in~\cite{NeherShraiman2009},
which are stochastic differential equations for the frequencies and pair-wise correlations
($f_i(\alpha)$ and $M_{ij}(\alpha,\beta)$ in our case).
When translated  
to equations for $h_{i}(\alpha)$ and $J_{i,j}(\alpha,\beta)$ near fixation
the noise will be large, which destabilizes \eqref{eq:final-result}.

\section{DCA for whole-genome sequencing data}
\label{sec:DCA-sparse}
\begin{figure}
\centering
\includegraphics[width=0.95\linewidth]{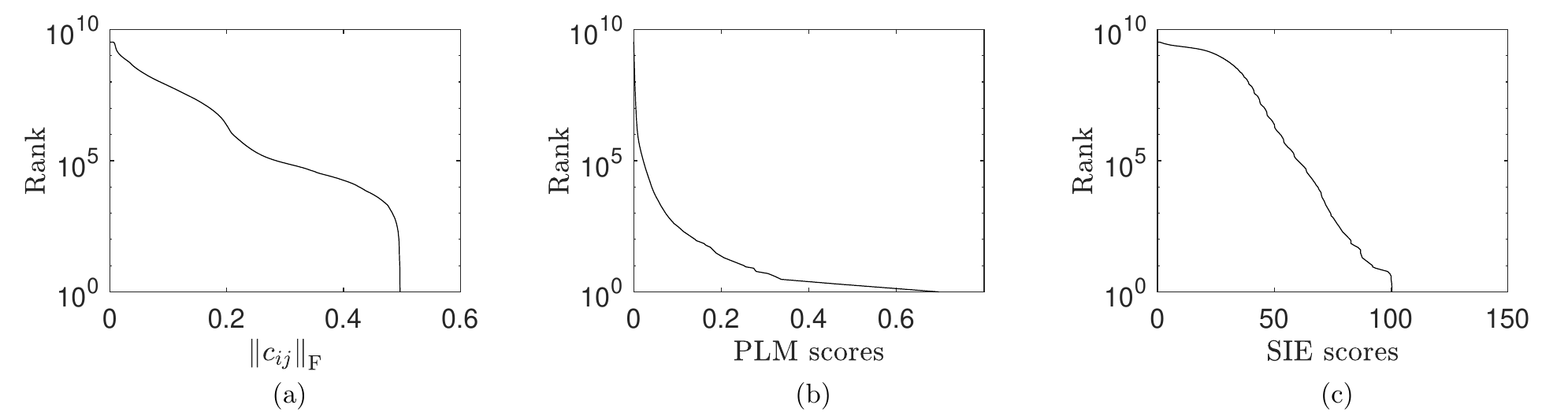}
\caption{Cumulative distributions of (a) correlations; (b) pseudo-likelihood maximization (PLM); (c) small-interaction expansion (SIE); 
Semi-logarithmic scale visualizing the distributions of of approximately $10^{10}$ elements. 
The scalar value associated to each pair of loci $i$ and $j$ is the Frobenius norm
of the $3\times 3$-correlation matrix (case a) or the Frobenius norm of the inferred 
Potts model $3\times 3$-matrix element (cases b and c). 
}
\label{fig:distributions-accumulated}
\end{figure}

\begin{figure}
\centering
\includegraphics[width=0.95\linewidth]{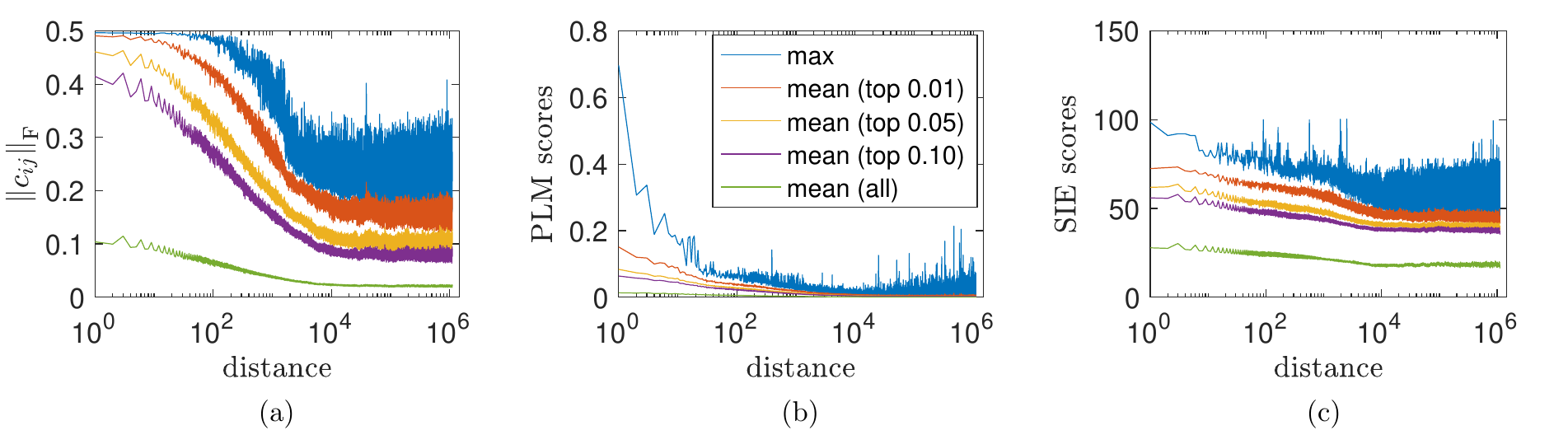}
\caption{Distributions as functions of genomic distance: (a) correlations; 
(b) pseudo-likelihood maximization (PLM).
(c) small-interaction expansion (SIE).
Same data and same norms as in Fig.~\protect\ref{fig:distributions-accumulated}.
Figures show averages in windows of genomic distance.
Blue: maximum, red: top-$1\%$, yellow: top-$5\%$, violet: top-$10\%$, green: mean, all in window.
The curves in (b) show a sharp initial decrease with genomic distance which generally much lower
values beyond genomic distance $10^3$ where recombination can be expected to act effectively.
}
\label{fig:distributions-genomic-distance}
\end{figure}

A set of whole-genome sequences of the human pathogen \textit{S. pneumoniae} obtained in the Maela collaboration (Materials \& Methods) 
can be represented as about $3,000$ genotypes 
of about $100,000$ loci each. 
Correlations and Potts model terms obtained from this data have qualitatively different distributions,
as shown in Fig~\ref{fig:distributions-accumulated} and Fig~\ref{fig:distributions-genomic-distance}. 

The number of correlations larger than a cut-off $c$ grows quickly when $c$
decreases below its maximum value,
while the cumulative distribution of inferred Potts model couplings
have a much more pronounced tail.
This implies that the set of largest DCA couplings is better separated than the largest correlations from the unavoidable
background due to under-sampling~\cite{Xu-2017a,Xu2018}.
Correlations also generally have a more uniform distribution
across genomic distance, 
while the representation as a Potts model is more sparse (Fig~\ref{fig:distributions-genomic-distance} (a) and (b)),.
The first-order perturbative version of DCA employed in~\cite{NeherShraiman2009,NeherShraiman2011}
here called SIE
gives in both instances results closer to correlations, \textit{see} Fig~\ref{fig:distributions-accumulated} (c) and Fig~\ref{fig:distributions-genomic-distance} (c).

\begin{figure}
\centering
\includegraphics[width=0.95\linewidth]{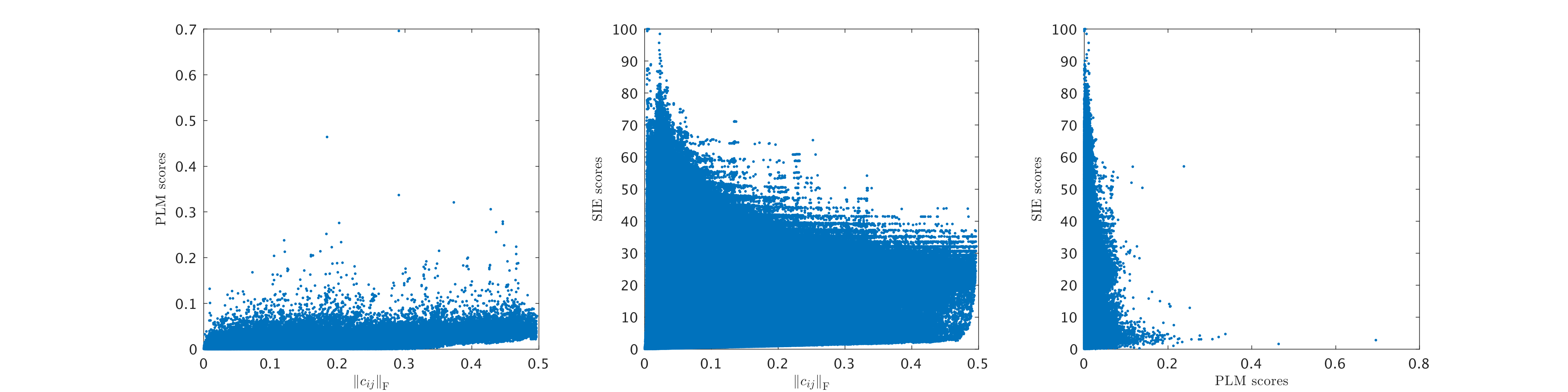}
\caption{Pair-wise scatter plot of correlations, SIE and PLM: 
(a) PLM vs correlations; 
(b) SIE vs PLM;
(c) SIE vs PLM.
Same data and same norms as in 
Fig~\protect\ref{fig:distributions-accumulated} and Fig~\protect\ref{fig:distributions-genomic-distance}. 
The numerical scale in each direction depends on the 
details of the norms and inference procedure, \textit{e.g}
$PLM$ scores depend on $L_2$ regularization parameters.
Correlations and PLM scores are numerically similar while SIE scores are not, as discussed in text.
}
\label{fig:Maela-scatter-plot}
\end{figure}

Fig~\ref{fig:Maela-scatter-plot} shows pair-wise scatter-plots of correlations and DCA terms
obtained by PLM and SIE.
In all three cases the scatter-plots are ``clouds of points'', indicating that DCA and correlations
measure different properties of the data. 
The scatter-plot of PLM vs correlations shows a weak trend,
such that larger PLM scores are associated to larger correlations.
Such trends are absent in correlations-SIE and PLM-SIE, and
SIE values are also numerically large.
In fact, the correlation matrix is under-sampled, hence smaller correlations reflect sampling noise 
and this is also an issue for SIE, as well as the sensitive dependence on almost fixed alleles in this procedure.
PLM scores and correlations were compared graphically for this data in~\cite{GaoZhouAurell2018}, 
with a cut-off excluding short-range interactions.

\section{The QLE phase is obscured by genetic drift}
\label{sec:QLE-genetic-drift}
\begin{figure}
\includegraphics[width=.95\linewidth]{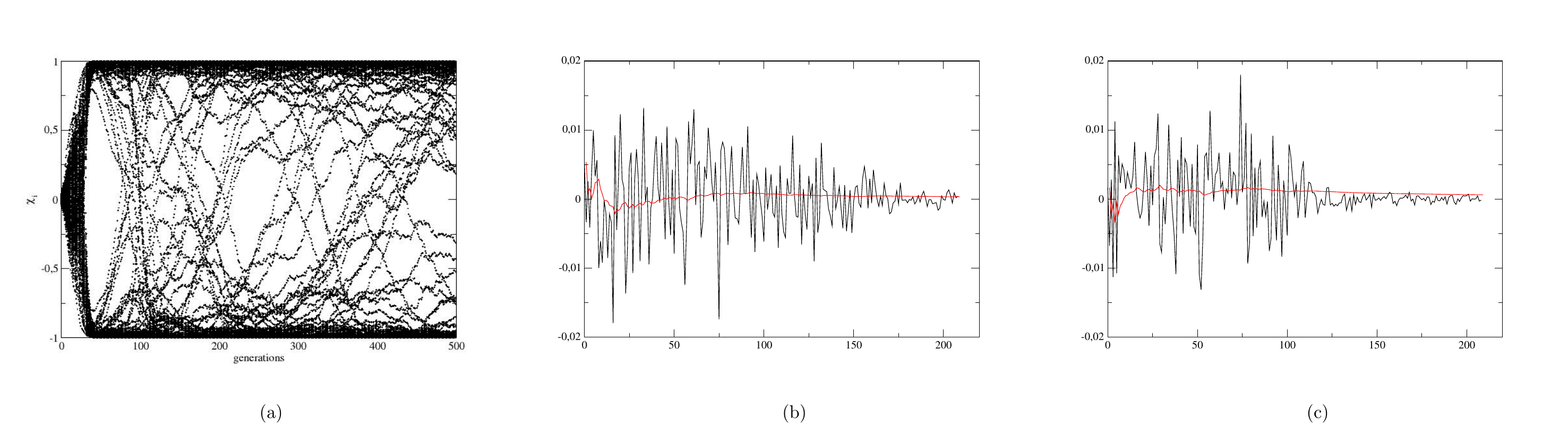}
\caption{\label{fig:time-development}  
Temporal behavior of (a) all magnetizations defined as $\chi_i=f_i(2)-f_i(1)$
and ((b) and (c)) two selected correlations defined as $\chi_{ij}=f_{ij}(1,1)-f_{ij}(1,2)-f_{ij}(2,1)+f_{ij}(2,2)$ 
in a simulation of a population of Ising genomes (two alleles per locus). 
Number of loci $L$ of the genotypes is 256, number of genotypes $N$ in the population is 50000,
other simulation parameters as reported in Materials \& Methods.
Data is taken every five generations, total simulation time is $2500$ generations.
}
\end{figure}

Genetic drift is the random changes from one generation to the next due to chance events.
In a finite population statistical genetics as described above only holds on the 
average; when following one population in time fluctuations of order $N^{-\frac{1}{2}}$ appear
for observables such as single-locus frequencies and pair-wise loci-loci correlations.
Fig.~\ref{fig:time-development} 
reports simulations using the FFPopSim software showing that these fluctuations can in practice be quite large, even 
for populations that are not small.

According to the theory developed 
in~\cite{NeherShraiman2011} (Appendix C)
dynamics of correlations is relaxational and the curves of correlations
vs time hence should fluctuate around an equilibrium value, which 
is the one given in \eqref{eq:main-Neher-Shraiman-eq} above.
The fluctuations in Fig.~\ref{fig:time-development} are however large compared to the pair-wise fitness values,
and DCA inference from instantaneous values of the ensemble correlations
are not good predictors for pairwise fitness (data not shown).
The dynamics of frequencies is not relaxational, and one may hence observe
large changes where the population at one locus changes from one allele to another.
A further conclusion is that SIE should not here be an appropriate 
inference procedure also because fluctuations 
in the frequencies are large and have long time scales; flavors of DCA
that rely only on correlations should exhibit better performance.

\section{Discussion}
\label{sec:discussion}
The main question addressed in this work is if and when DCA can be expected to work
for genome-scale epistasis analysis. We have given an answer in the context
of statistical genetics: for a population evolving under 
recombination, mutations and fitness variations this is so when 
recombination is sufficiently fast. The joint distribution of the population over genotypes
then approximately factorizes into a product of identical Potts distributions (\ref{eq:Potts}). 
Treating a set of genomes as independent samples from such a distribution
allows to infer fitness parameters ($F_{ij}$) from Potts model parameters ($J_{ij}$)
by inverting \eqref{eq:main-Neher-Shraiman-eq}, and this is essentially what using DCA on such data means.
We now discuss limits to the analysis and further directions. 

The first limitation is that 
DCA cannot be expected to yield meaningful results when recombination is weak.
One example of such an effect was already given in~\cite{Skwark-2017a} 
where also data from \textit{Streptococcus pyogenes} was presented (Fig. 6 of~\cite{Skwark-2017a}).
Another example was recently given on
\textit{Vibrio parahaemolyticus}, a human gastrointestinal pathogen
in panmixia \textit{i.e.} where all strains are able to recombine, but  
having a very low overall recombination rate~\cite{Falush2018}.
The problem of inferring fitness from data on populations that are in the CC phase 
appears to be both conceptually and practically important; 
we hope to be able to return to such questions in future work.

A second limitation concerns finite populations, particularly simulated data, where the population has to be 
of moderate size.
According to the theory developed for Ising genomes in~\cite{NeherShraiman2011} (Appendix~C)
and qualitatively confirmed above in Fig.~\ref{fig:time-development},  
frequencies and correlations follow stochastic differential equations 
with noise strength scaling as $N^{-\frac{1}{2}}$. 
In principle Potts model parameters ($h_i(\alpha)$ and $J_{ij}(\alpha,\beta)$)
for categorical data also follow stochastic differential equations, but of a more complicated form
due to the inverse Ising/Potts relations.
Applying the DCA procedure to finite-$N$ data thus requires 
parameter inference from a high-dimensional stochastic time series
with a complicated deterministic part. This may not be an easy task.

A third limitation is the neglect of spatial and environmental separation. 
Bacteria such as the human pathogen~\textit{Helicobacter pylori} readily recombine if they meet, but can only do so when
their human host populations overlap~\cite{Thorell2017}. 
Allele frequencies may be different for different bacterial populations, reflecting
differences in the host populations and environments. 
If data from the different populations is pooled, this will be a confounding factor for some flavors of DCA 
\textit{e.g.} for PLM and SIE, 
These types of issues appear to merit further study.  

\section*{Acknowledgments}
E.A. thanks Prof Boris Shraiman for enlightening discussions.
This research was supported by National Science Foundation of China (grant numbers 
11647601 and 11421063). The numerical computations were partly carried out 
at the HPC cluster of ITP-CAS.

\appendix

\section{The small-interaction expansion (SIE) for categorical data}
\label{app:SIE}
We need the
solution of the the matrix equation $\mathbf{u}_{\alpha}=\sum_{\beta}M_{ii}(\alpha,\beta) \mathbf{v}_{\beta}$ in the space of vectors orthogonal 
to $(1,1,\ldots,1)$, where the one-locus allele correlation matrix $M_{ii}(\alpha,\beta)$ is defined in \eqref{eq:c-ii}. That is given by
\begin{equation}
    \label{eq:c-inverse-operation}
  v_{\alpha} = \frac{u_{\alpha}}{f_i(\alpha)} - \frac{1}{M_i}\sum_{\beta=1}^{M_i} \frac{u_{\beta}}{f_i(\beta)}
\end{equation}

\begin{figure}
\includegraphics[width=.95\linewidth]{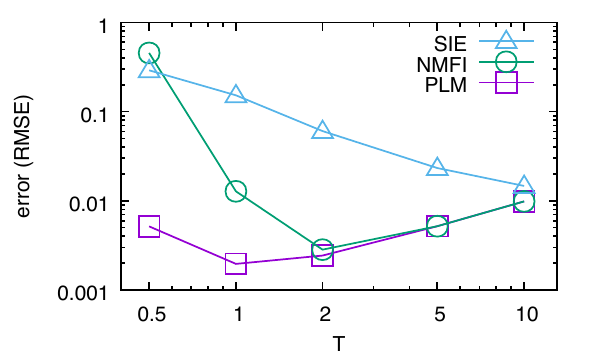}
\caption{\label{fig:T-rmse-J}  
Root mean square error of three inference algorithms on Sherrington-Kirkpatrick (SK) spin glass.
Abscissa ($x$-axis): temperature.
Ordinate ($y$-axis): pseudo-likelihood maximization (PLM),
naive mean-field (NMFI) and small-interaction expansion (SIE).
The SK model is a widely studied test case
in the DCA literature; performance of other inference algorithms
can be found in \textit{cf}~\protect\cite{Nguyen-2017a}, and references cited therein.
} 
\end{figure}

Consider now the Potts model when all the interaction parameters $J_{ij}(\alpha,\beta)$ are small.
One frequency can be estimated to zeroth and first order 
as
\begin{eqnarray}
  f_i(\alpha) &=& \frac{e^{h_{i}(\alpha)}}{N_i} +\sum_{j,\beta} J_{i,j}(\alpha,\beta) \frac{e^{h_{i}(\alpha)}}{N_i}  \frac{e^{h_{j}(\beta)}}{N_j} -\nonumber \\
  && \sum_{j,\beta,\gamma} J_{i,j}(\gamma,\beta) \frac{e^{h_{i}(\alpha)}}{N_i} \frac{e^{h_{i}(\gamma)}}{N_i} \frac{e^{h_{j}(\beta)}}{N_j}
\end{eqnarray}
where we have used the abbreviation $N_i = \sum_{\alpha} e^{h_{i}(\alpha)}$. The fluctuation-dissipation relations for the Potts model read
\begin{equation}
  M_{ij}(\alpha,\beta) = \frac{\partial f_i(\alpha)}{\partial h_{j}(\beta)}
\end{equation}
and therefore, comparing (\ref{eq:c-ii}),
\begin{equation}
  \label{eq:Neher-Shraiman-Potts}
  M_{ij}(\alpha,\beta) = \sum_{\gamma,\delta}  J_{i,j}(\gamma,\delta) \, M_{ii}(\alpha,\gamma) \, M_{jj}(\beta,\delta)
\end{equation}
Since the Potts parameters are in the Ising gauge~\cite{Weigt-2009a,Ekeberg-2014a} the matrix multiplications 
in \eqref{eq:Neher-Shraiman-Potts}
can be inverted using \eqref{eq:c-inverse-operation}:
\begin{eqnarray}
  J_{i,j}(\alpha,\beta) &=& \frac{ M_{ij}(\alpha,\beta)}{f_i(\alpha)f_j(\beta)} - \frac{1}{f_j(\beta)} \sum_{\gamma}  \frac{M_{ij}(\gamma,\beta)}{f_i(\gamma)}  \nonumber \\
    \label{eq:Neher-Shraiman-Potts-inverse}
                     && - \frac{1}{f_i(\alpha)}\sum_{\delta}  \frac{M_{ij}(\alpha,\delta)}{f_j(\delta)} + \sum_{\gamma,\delta}  \frac{M_{ij}(\gamma,\delta)}{f_i(\gamma)f_j(\delta)}
\end{eqnarray}
This can be interpreted as an inference algorithm where the interaction parameters $J_{ij}(\alpha,\beta)$ are
determined from the single-locus frequencies $f_i(\alpha)$ and pair-wise locus-locus correlations $M_{ij}(\alpha,\beta)$.
In this paper we refer to this method, introduced for Boolean data in~\cite{NeherShraiman2009,NeherShraiman2011}, as \textit{SIE}.
Since it is a first-order perturbative solution to naive mean-field inference, SIE can
be expected to be a comparatively weak inference procedure. Fig~\ref{fig:T-rmse-J} confirms that this is the case
for the Sherrrington-Kirkpatrick spin glass model.  
An implementation of SIE for categorical data can be found at~\url{github.com/gaochenyi/DCA-QLE}.

\section{\textit{S. pneumoniae} sequence data} 
\label{app:sequence-data}
Whole-genome sequences of carriage isolates
from two birth cohorts of infants and their mothers in the Maela refugee 
camp (Thailand)~\cite{Turner2012,Turner2013}
were reported in~\cite{Chewapreecha2014}.
This data  
was filtered for positions (loci)
that carry at most two alleles and a moderate amount of gaps, as described previously~\cite{Skwark-2017a,GaoZhouAurell2018}.
This procedure results in $3,145$ genotypes each containing $81,506$ loci, where the alleles at each locus can take
three values (major, minor, gap).
The original MSA data can be found in~\cite{Skwark-2017a}, while the filtered MSA can be 
retrieved by the pipeline function in~\url{github.com/gaochenyi/DCA-QLE}.

\section{Correlation matrix computations}
\label{app:correlations}
Correlation matrices were computed 
using the MATLAB implementation 
available at~\cite{Gao-github} (\url{github.com/gaochenyi/CC-PLM}).
On the Maela data set ($L=10^5$) the compute time was approximately
30 core-hours using a 
56-core server with four Intel Xeon E7-4850~v3 processors.
The run-time memory used is about $70$~GB storing all correlations in memory.

\section{Direct Coupling Analyses}
\label{app:DCA-methods}
Potts model parameters were inferred by
the asymmetric
$\ell_2$-regularized pseudo-likelihood maximization 
method~\cite{Ekeberg-2014a} using the software PLM
at~\cite{Gao-github} 
(\url{github.com/gaochenyi/CC-PLM}).
On the same data set and in the same compute environment
as above, the total compute time was about $20,000$ core-hours.
The implementation of naive mean-field inference (NMFI) for categorical data
used in the paper can be found at~\url{github.com/gaochenyi/DCA-QLE}.

\section{Simulations of Wright-Fisher model with pairwise fitness function and recombination}
\label{app:simulations}
Simulations of the Wright-Fisher model with recombination
were done with the FFPopSim simulation package~\cite{FFPopSim}
with parameter settings as given in Table~\ref{fig:FFPopSim-parameters}.

\begin{table}
\centering
\caption{Parameters for FFPopSim simulations}
\begin{tabular}{lr}
number of loci (L) & 100 \\
circular & False \\
number of traits & 1 \\
population size & 126641 \\
carrying capacity ($N$) & 500 \\
generation & 0 \\
outcrossing rate ($r$) & 1.0 \\
crossover rate ($\rho$) & 0.05 \\
recombination model & CROSSOVERS \\
mutation rate ($\mu$) & 0.01 \\
participation ratio $Y$ & 0.0022 \\
number of non-empty clones ($N$) & 500 \\
\end{tabular}
\label{fig:FFPopSim-parameters}
\caption{Parameter settings for the simulations
reported as Figs.~\protect\ref{fig:time-development} in main text.}
\end{table}
In the simulations reported in Fig.~\ref{fig:time-development} 
single-locus contributions to fitness (parameters $F_i$) are zero, 
while pair-wise loci-loci contributions (parameters $F_{ij}$) are random and small ($\pm 7.8\times 10^{-5}$).

\bibliography{ref,ref2}

\end{document}